\documentstyle[11pt,newpasp,twoside,epsf]{article}
\def\edcomment#1{\iffalse\marginpar{\raggedright\sl#1\/}\else\relax\fi}
\reversemarginpar

\begin{document}
\title{The BMW Deep X--ray Cluster Survey}
\author{Alberto Moretti, Luigi Guzzo, Sergio Campana, Stefano Covino, 
Davide Lazzati, Marcella Longhetti, Emilio Molinari, Maria Rosa Panzera, 
Gianpiero Tagliaferri}
\affil{Osservatorio Astronomico di Brera, Via E. Bianchi 46, I-23807 Merate, Italy}
\author{Ian Dell'Antonio}
\affil{Physics Department, Brown University, Box 1843,
Providence, RI, USA}
\begin{abstract}
We describe the main features of the BMW survey of serendipitous X--ray
clusters, based on the still unexploited ROSAT--HRI archival observations.
The sky coverage, surface density and first deep optical CCD images of the
candidates indicate that this sample can represent an excellent complement
to the existing PSPC deep cluster surveys and will provide us with a
fully independent probe of the evolution of the cluster abundance, in
addition to significantly increasing the number of clusters known at $z>0.6$.
\end{abstract}

\section{Introduction}
In the last few years, X--ray selected samples of clusters of galaxies
have become a formidable tool for cosmology.  Deep surveys using ROSAT
PSPC archival data have been used to study the evolution of the
cluster abundance and X--ray luminosity function (XLF) and constrain
cosmological parameters (e.g. Borgani et al. 1999).  The lack of
evolution of the XLF observed for $L \sim L^*\simeq 4\cdot
10^{44}\,h^{2}$ erg s$^{-1}$ out to $z\sim 0.8$ favours low values for
$\Omega_M$ under reasonable assumptions about the evolution of the
$L-T$ relation.  At the same time, the original hint from the EMSS
(Gioia et al 1990; Henry et al. 1992) of evolution at the very bright
end of the XLF seems to be confirmed (Vikhlinin et al. 1998, Nichol et
al. 1999, Rosati et al. 2000 and references therein).  The main
statistical limitation of this conclusion rests with the small sky
coverage of the ROSAT deep surveys, which clashes with the intrinsic
rarity of highly luminous clusters.  XMM--Newton and Chandra are
already attracting justified attention as the likely source for future
samples, but no significant sets of serendipitously selected clusters
can be reasonably expected from these observatories for at least
another 2 years.  This presents the window of opportunity for our
survey, which uses data from the ROSAT High--Resolution Imager (HRI)
archive.  With respect to the PSPC, the HRI offers superior angular
resolution.  Our results indicate that it is actually a surprisingly
good source of samples of high-redshift clusters.

\begin{figure}
\centering
\plottwo{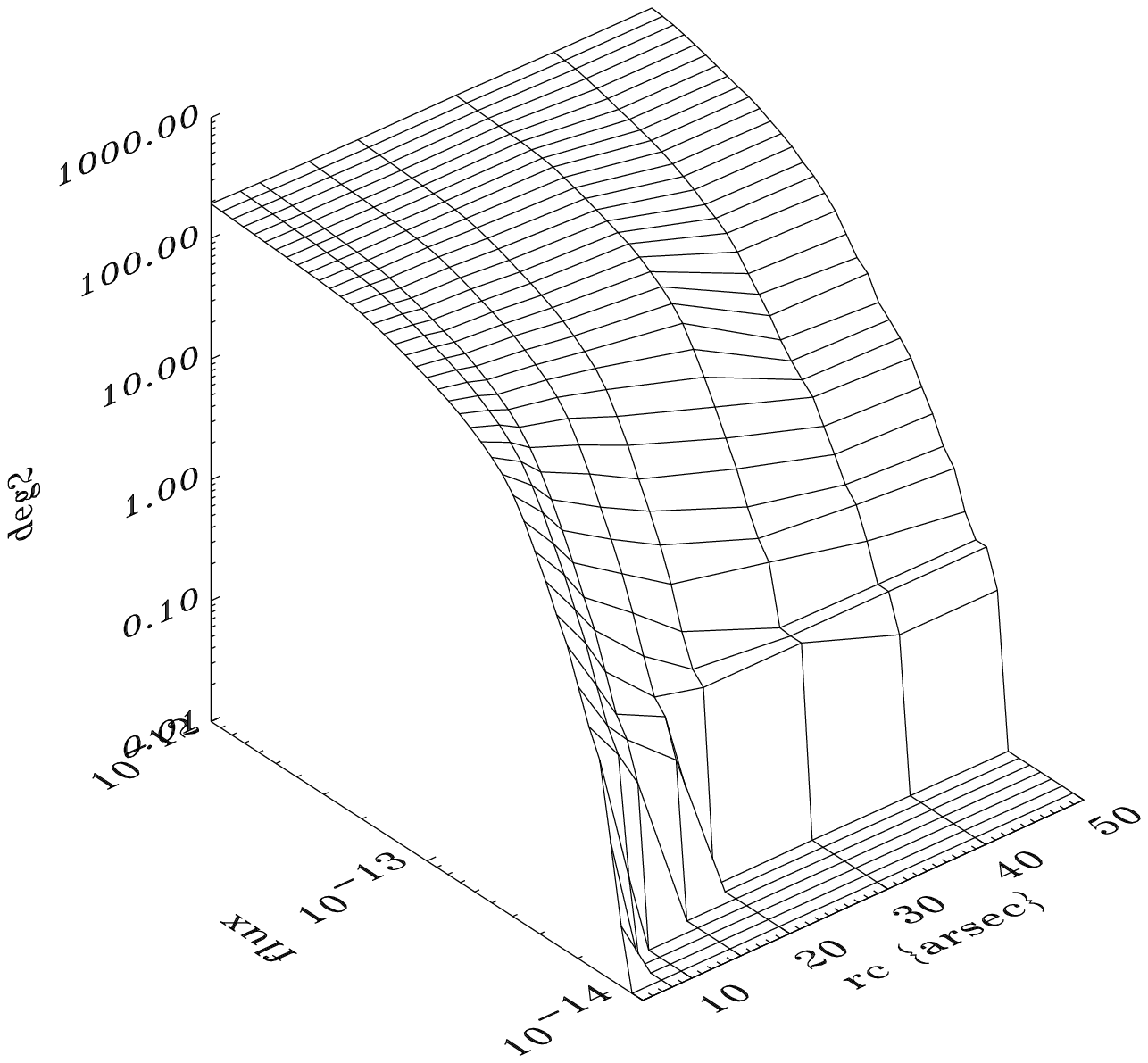}{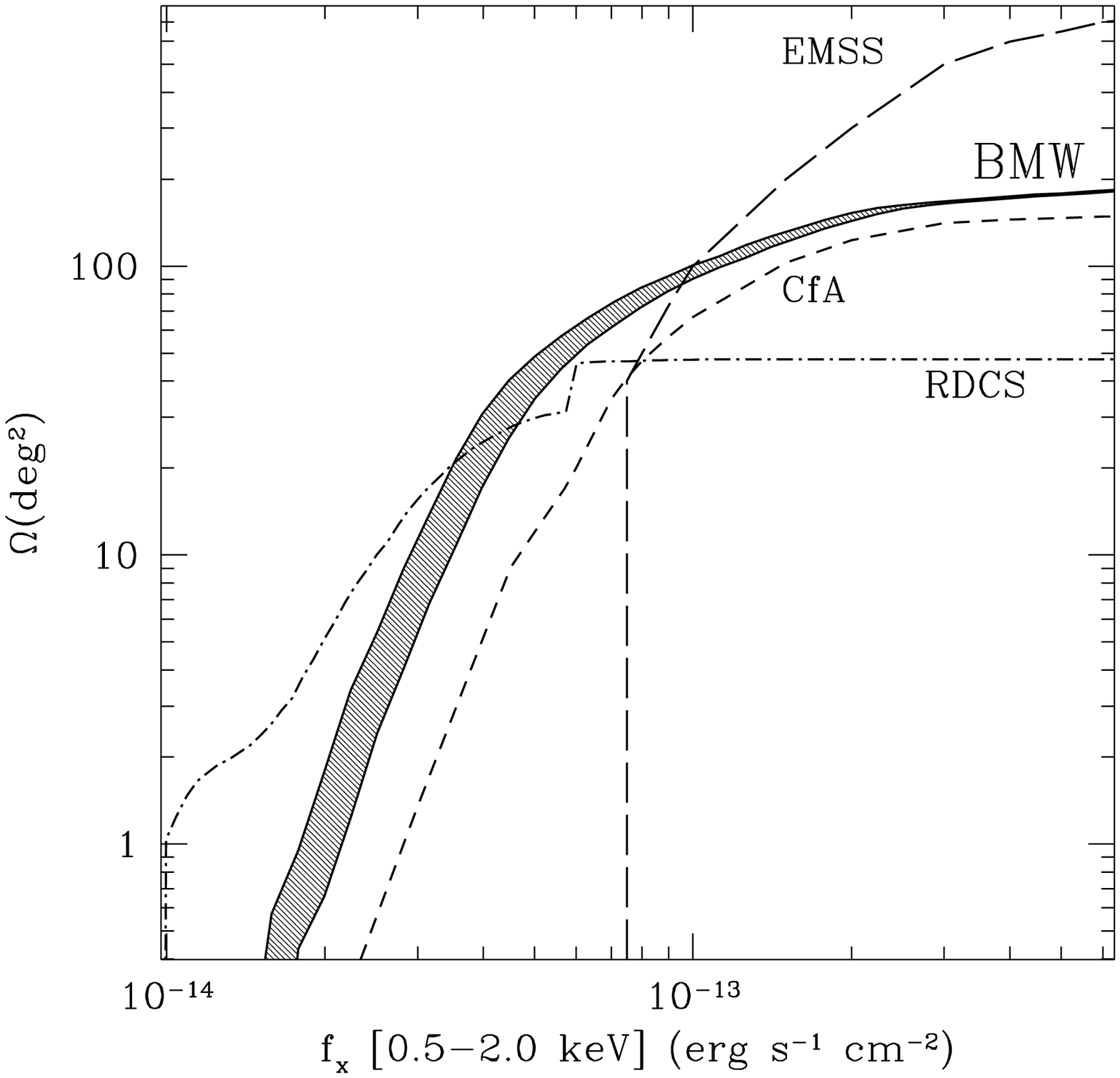} 
\caption[]{ {\bf Left:} Sky coverage of the BMW survey, as a function
of X--ray flux and extension of the sources.  {\bf Right:} Same, but
for a typical faint--source core radius $\sim 10$ arcsec, compared to
some previous X--ray cluster surveys (see Rosati et al. 2000 for
relevant references).  Note the good compromise between the fairly
large area at intermediate fluxes ($\sim 100$ sq. deg. around $\sim
10^{-13}$ erg cm$^{-2}$ s$^{-1}$), and the depth of the BMW sample 
(1 sq. deg. at $2.5 \times 10^{-14}$ erg cm$^{-2}$ s$^{-1}$).} 
\label{fig:skyc}
\end{figure}

Our new X--ray selected sample of candidate clusters of galaxies
is based on the recently completed BMW survey of 
serendipitous X--ray sources over 3000 ROSAT HRI fields. The 
sample includes 287 candidates,
with a significantly large sky coverage in comparison with other
recent deep surveys.
We are conducting a multi-site imaging campaing to fully
identify the cluster sample.  First results at ESO and TNG are extremely
encouraging: with approximately an 80\% rate of photometric confirmation
in the first subsample of 35 candidates.

\section{The BMW Project}
The Brera Multi-scale Wavelet (BMW) project has currently completed the
systematic analysis of about 3100 HRI pointings using a wavelet
detection algorithm (Lazzati et al. 1999). This resulted in a catalog of
$\sim 19000$ serendipitous sources with measured fluxes and extensions
(Campana et al. 1999, Panzera et al. 2001). 
A clever selection of the HRI energy channels produced a reduction
of the background noise by a factor of $\sim 3$, thus greatly
improving the ability to detect low--surface--brightness sources as clusters.
The BMW general catalogue is built  excluding fields with $ |b_{II}| \le
20^{\circ} $ or pointed on the LMC and SMC.
Furthermore to build BMW cluster catalogue we have excluded cluster--targeted
HRI fields to avoid the bias produced by the cluster--cluster angular
correlation function, for which we have a clear positive detection in these
fields.
Cluster candidates were isolated on the basis of their extension,
selecting at a high significance level (corresponding to $>5 \sigma$)
and using only the well--sampled HRI area between 3 and 15 armin  off-axis.
We ended up with a list of 287 cluster candidates which were
visually classified on the DSS2 to reject obvious contaminants
(30 rejections, mostly nearby galaxies).
The BMW project is still under development such that a small assessment
in the absolute number is expected.

\section{Survey Sky Coverage}
An important parameter characterising a survey of serendipitous
sources as the BMW is its sky coverage, 
i.e. the effective solid angle covered
as a function of the limiting flux. 
Being based on archival pointings, different parts of the sky are
observed with different exposure times. 
Thus, for each 
value of the 
X--ray flux $f_x$ the sky coverage is 
given by
the 
total area of all
observed fields with limiting flux $f_{lim}\le f_x$.
In addition, due to the radial dependence of the point--spread
function (PSF) of the ROSAT X--ray telescope, within a single field the
effective flux limit is different for different off--axis angles and
extensions of the sources.  For this reason, the effective solid angle 
covered at different flux limits must be carefully estimated
considering the instrumental set--up and the detection and
characterisation methods adopted (e.g. Rosati 1995).  
We have therefore first estimated the sky coverage of the BMW survey for
extended objects by 
assuming a $\beta$ model with $\beta=2/3$, with a set of different core
radii. For each core radius we have convolved the analytic profile
with the PSF of the instrument at different off-axis angles and
we have calculated the corresponding maximum in the wavelet space.
Each HRI image has a detection threshold which is calculated in the wavelet
space (Lazzati et al. 1999) and is only function of the background.    
Thus for each image we 
could estimate directly the limiting detection flux
as a function of both the off-axis angle and source extension.
In Figure 1 we compare the BMW cluster sky coverage   
to some previous X--ray cluster surveys.
This 
relatively quick and straightforward method has some limitations
(e.g. it does not tell us how well the wavelet extension is measured), that 
need to be explored through simulations.  To this end, we are
currently running an extensive set of simulations 
following the approach of Vikhlinin  et al. (1998), and the first
outputs 
for a small sample of fields give results which are very close to the
semi--analytic calculation, confirming that the sky coverage of
Figure 1 should be a fair representation of our data.

\section{First Results from Optical Follow--up}

%
\begin{figure}
\plottwo {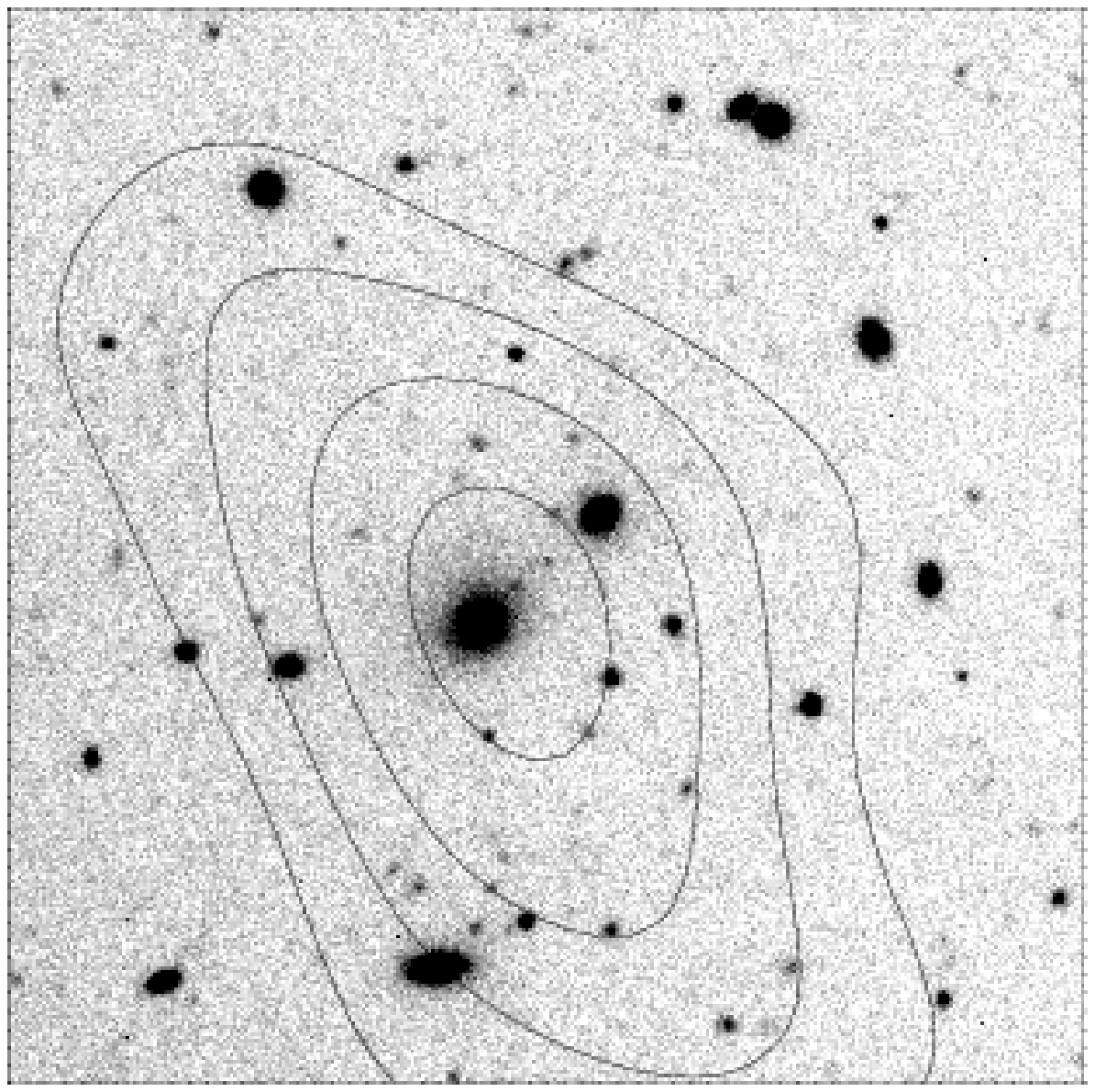}{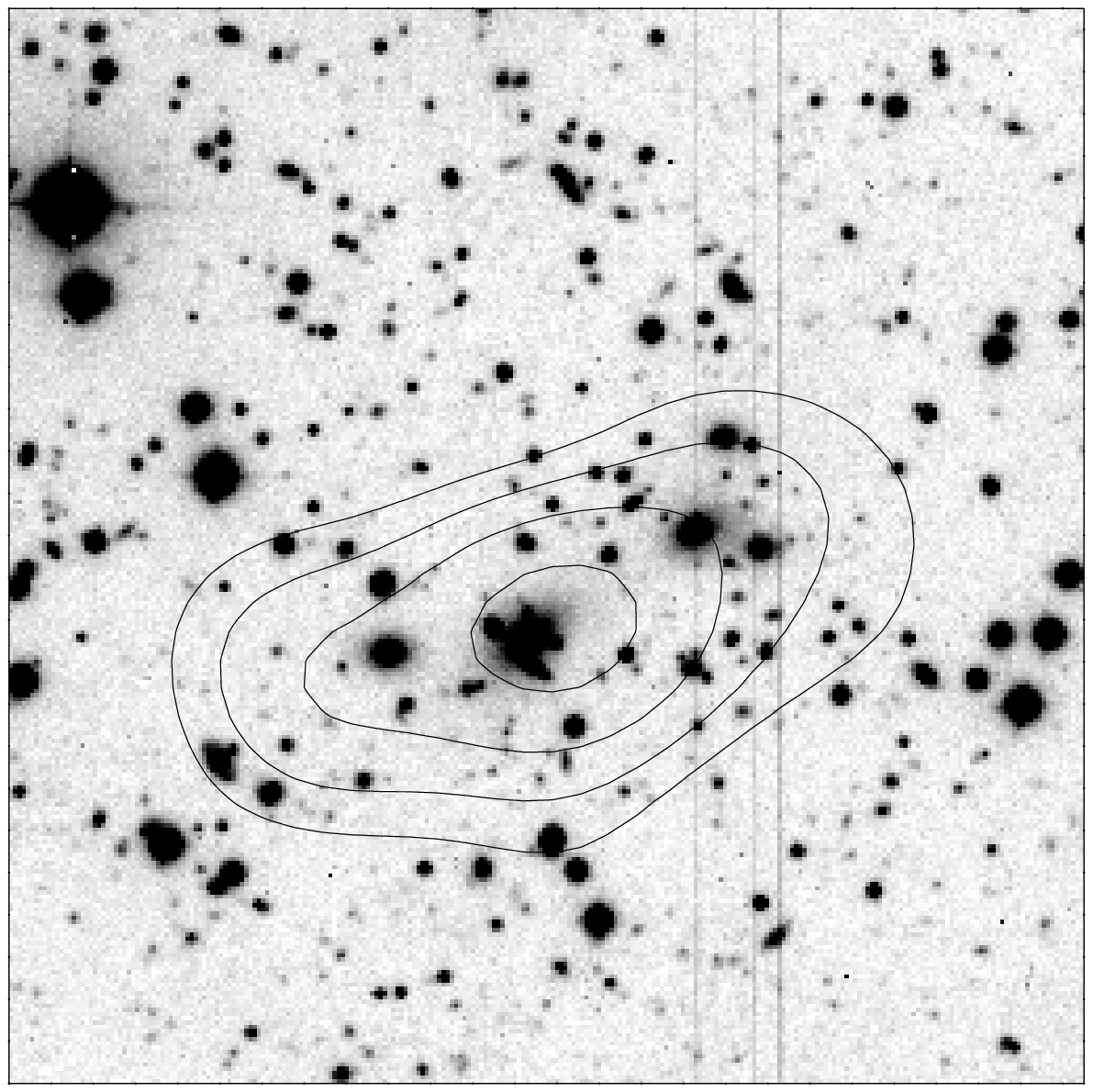}
\plottwo {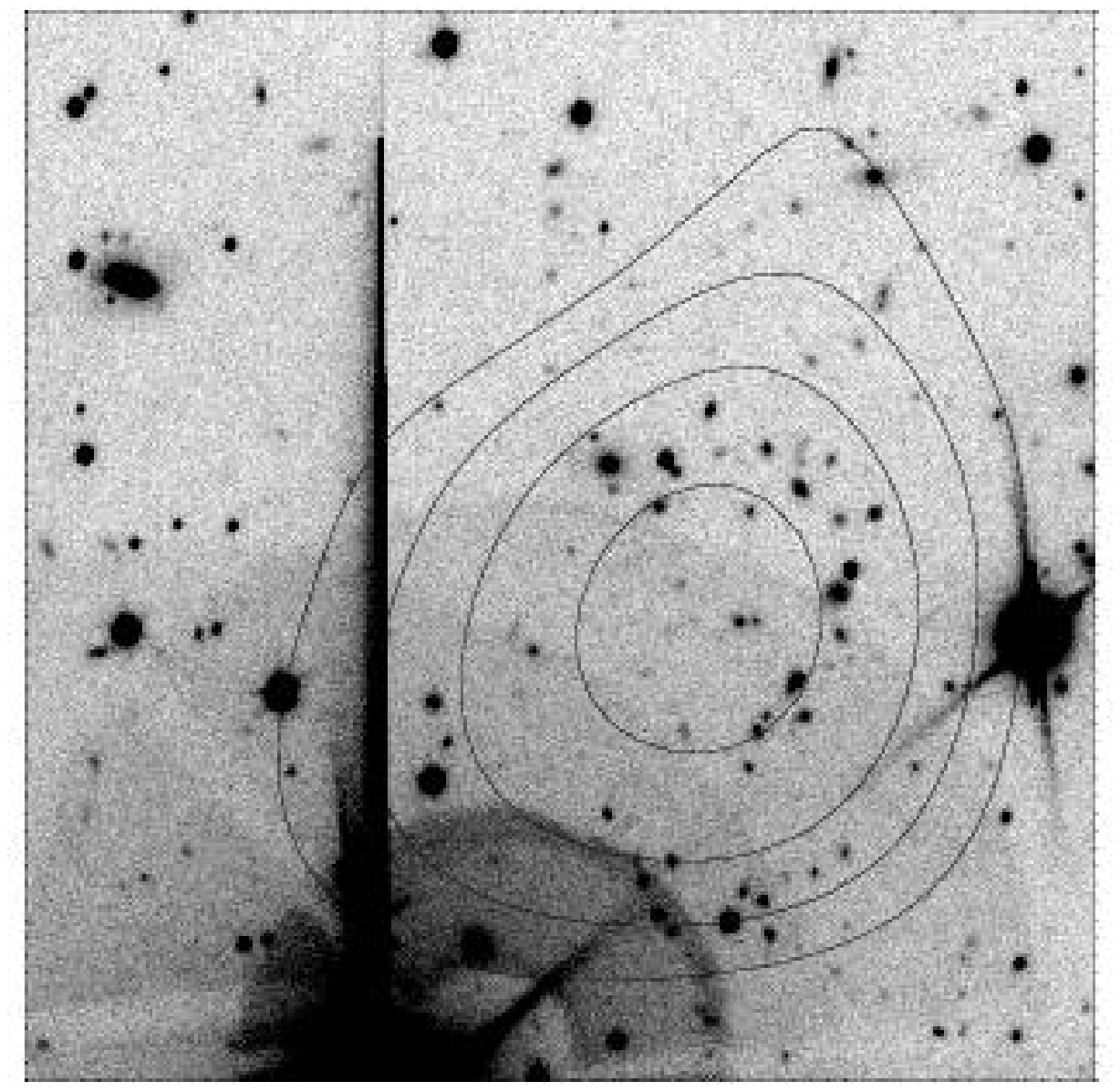}{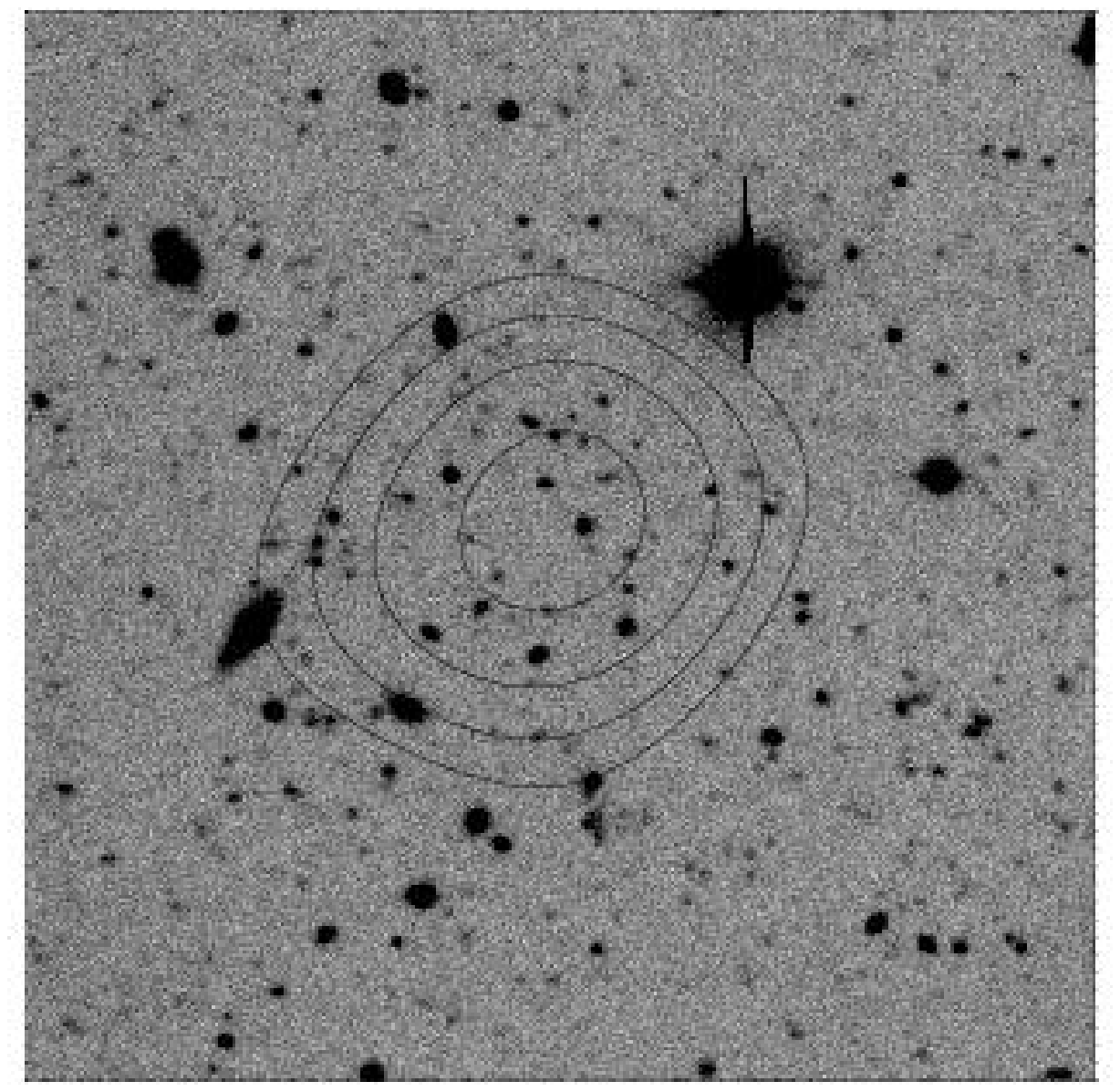}
\caption[]{Deep $g+r+i$ images (typical exposure times $\sim 5000$ sec) with
overlaid ROSAT--HRI X--ray contours, for a few examples of identified
groups/clusters in the BMW survey. Indicative redshifts for these
clusters range from $z\sim 0.2$ (top left) to $\sim 0.8$ (bottom right).
In each image the field of view is 3 x 3 arcmin.}
\label{fig:opt-X}
\end{figure}

\begin{figure}
\plottwo {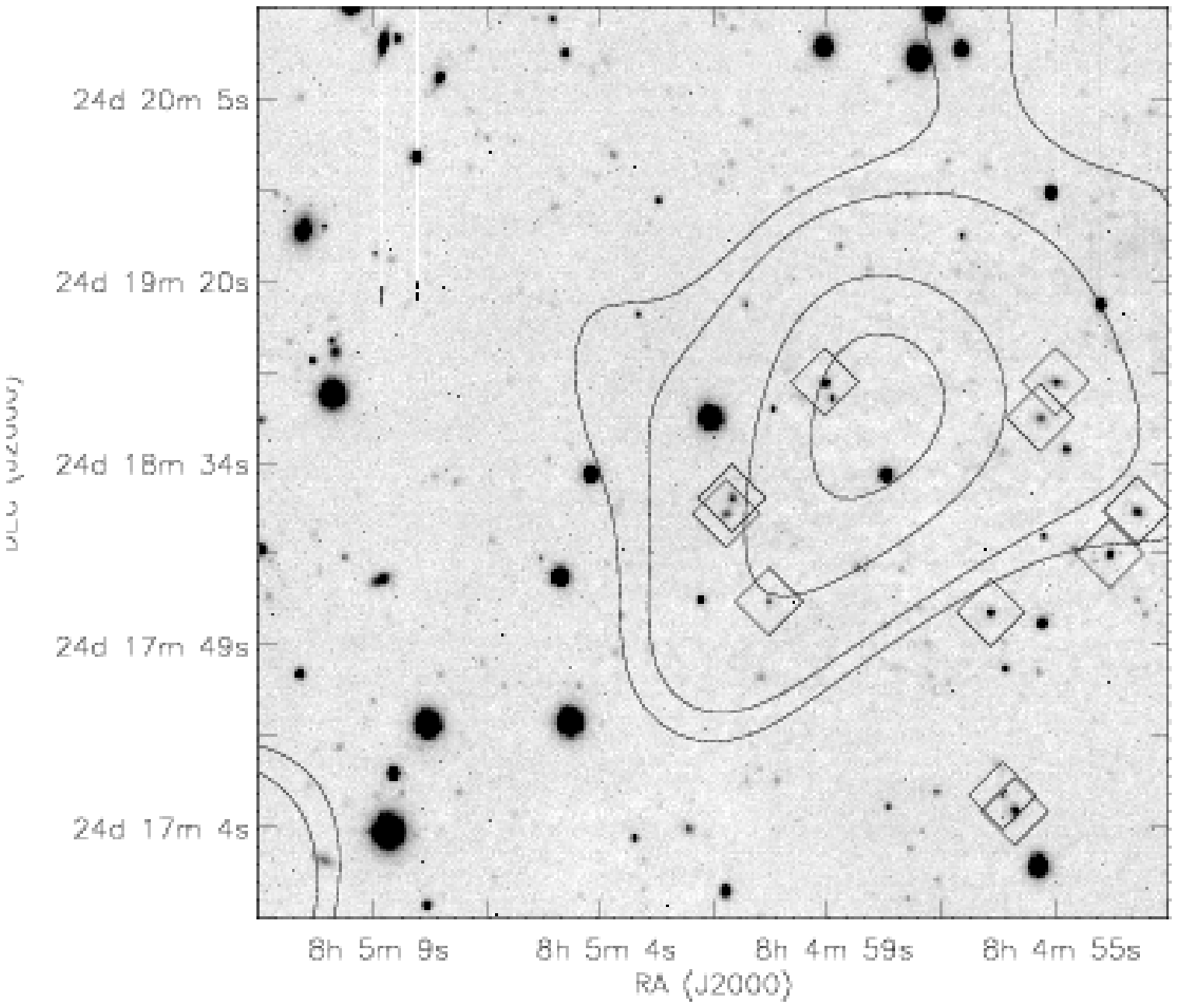}{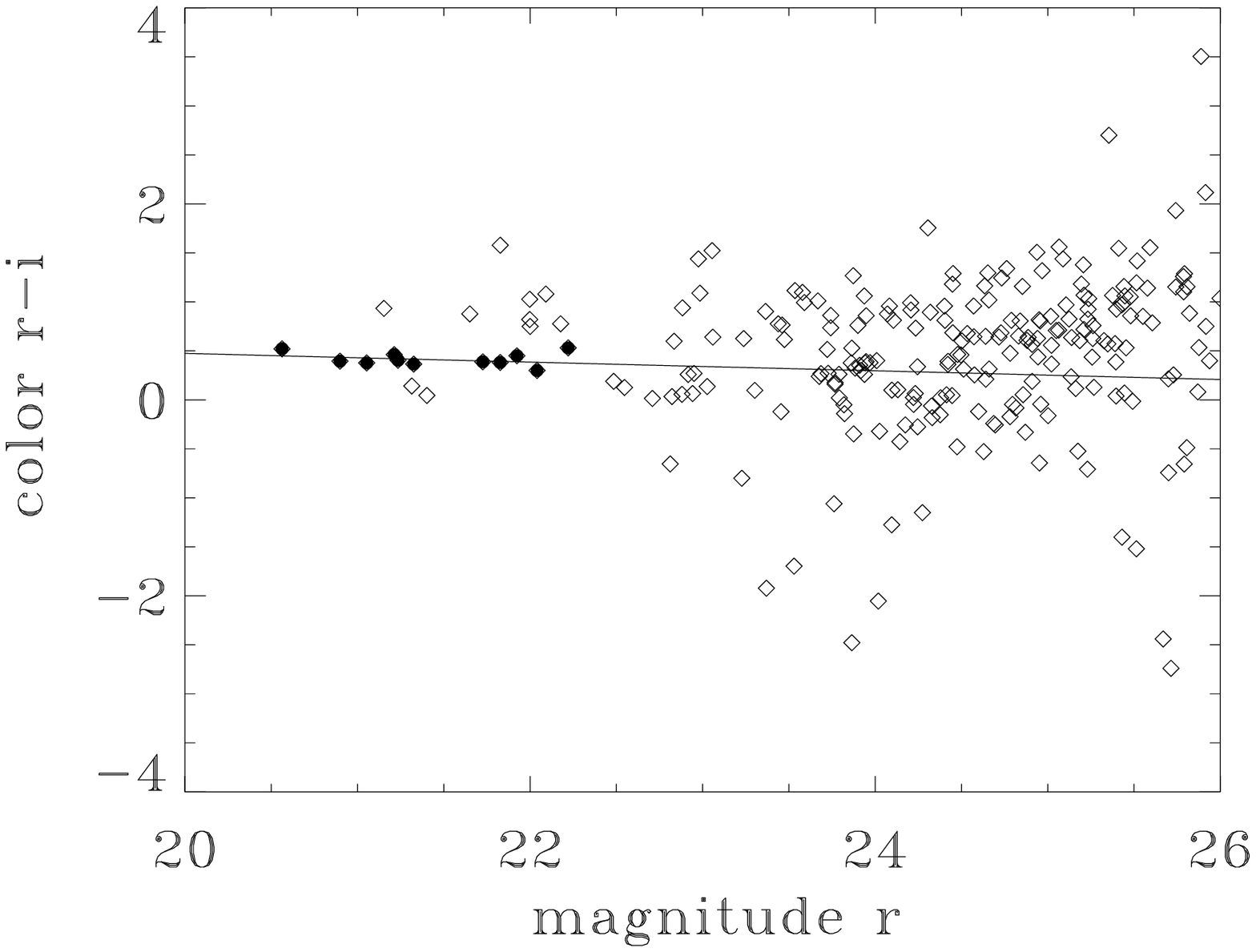}
\caption[]{CCD image of BMW 080459+241 ({\bf left}, 6000 sec exposure
in $r+i$ Gunn bands) with X--ray contours superimposed, and its
colour--magnitude diagram ({\bf right}).  A number of galaxies display 
a similar colour along an apparent {\sl red sequence} (filled dots)
and their position on the sky correlates significantly with the X--ray
emission (diamonds over left image).  The colour 
is that
expected for an early--type population at $z=0.6$.}
\label{tab:optfol}
\end{figure}

>From the 287 candidates we have further selected a {\it high priority} sample
of 165 objects by excluding the HRI fields with exposure time smaller then 10 ksec. 
In spring 2000, we started a long--term program of multi--band photometry
and spectroscopy of these fields, which is currently underway using telescopes
in both emispheres (mostly the TNG in La Palma and the ESO 3.6~m telescopes). 
We have recently (September 2000) reached a total of 35
candidates for which deep optical imaging has been secured in at least 
two bands. Preliminary analysis of these observations suggests a
success rate (i.e. evidence for a galaxy overdensity correlated with
the X--ray source) of about 80\%.  The still unidentified 20\%
fraction does not show any obvious pathology and we plan to add deep 
imaging in $K^\prime$ band, where the contrast of early--type galaxies 
is maximised, to definitely ascertain their nature. 

\section{Discussion}

A hint of the scientific potential of the BMW catalogue can be obtained
from the left panel of Figure 4 where the expected number
of clusters as a function of redshift is plotted.  These predictions use 
the computed BMW sky coverage of the {\it high priority} sample (165 objects)
and integrate the local X--ray luminosity function (De Grandi et al. 1999)
considering or not the evolution suggested by the RDCS, as reported in
Rosati (1999).
Simple comparison to the total expected numbers from Figure 4
would seem to imply that the BMW survey sees evolution in the XLF similar
to the RDCS results.
However, one has to await for the completion of the identification
campaign to be able to place serious constraints on evolution.
We should remark here the potential advantage of working with HRI data.
In the right panel of Figure 4 we have plotted the apparent 
diameters of a rich cluster ($ r_c=250$ kpc) and of a group of galaxies
 ($r_c=100$ kpc) as a function of redshift, together with a measure of
the 
resolution of the ROSAT HRI compared to the PSPC. As one can see, the
PSPC is not as suitable as the HRI for distinguishing groups of galaxies from
point-like sources beyond redshift 0.4.  Of course, this simplistic
plot depends on several variables, as the underlying cosmology, the
source profile and in particular does not take into account the
different noise level of the 2 instruments.  Nevertheless, it shows
that we should be able to explore the faint end of the XLF, the realm
of groups, which could not be studied from PSPC data.  In fact, the
potentially large number of clusters in the BMW sample, when compared
to prediction of evolution or no--evolution models, could also
be produced by a steeper local XLF.

\begin{figure}
\centering
\plottwo {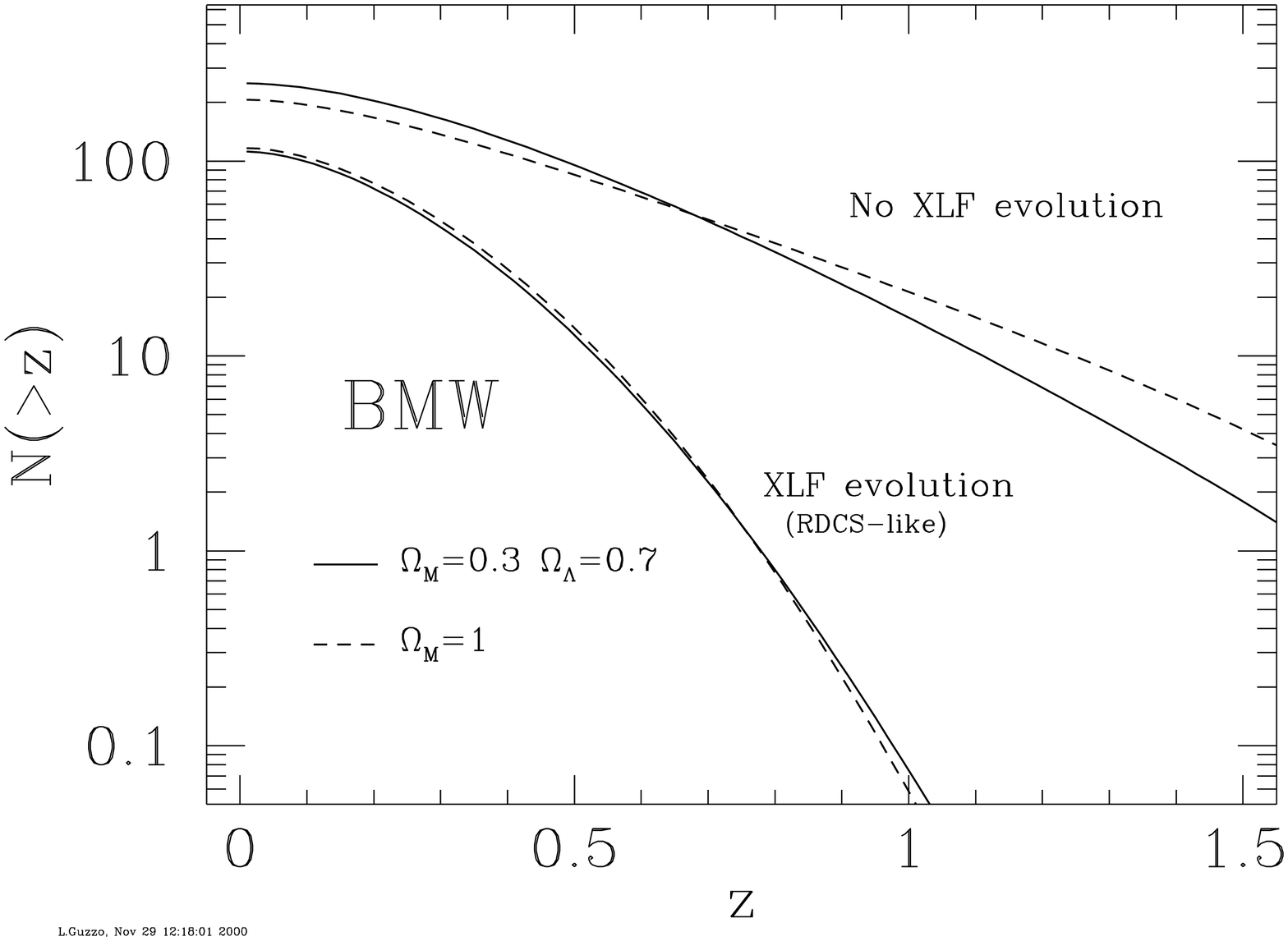} {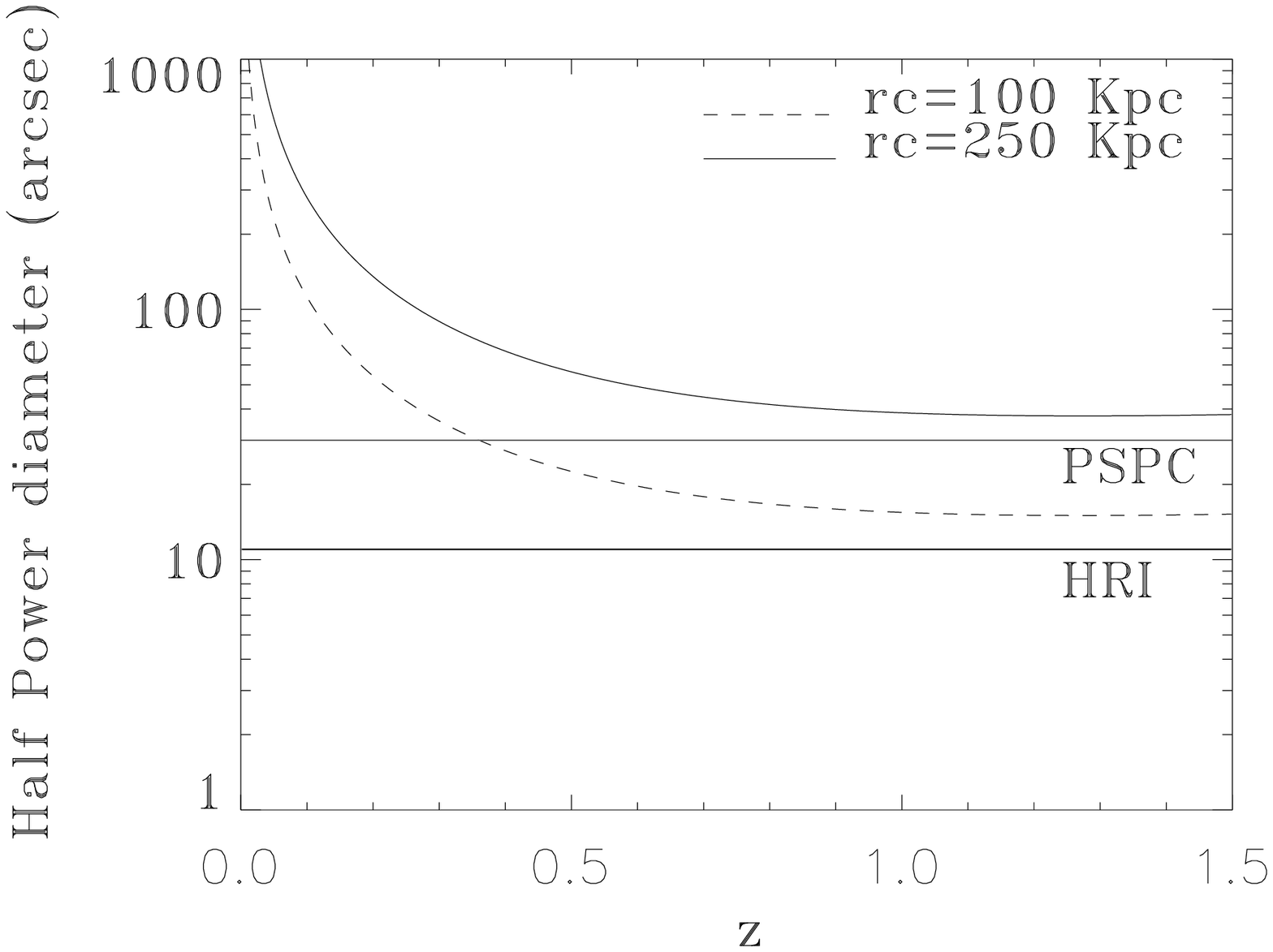}
\caption[]{ {\bf Left:} Expected integral distribution of BMW
clusters as a function of redshift, with or without the evolution of the
X--ray luminosity function suggested by the RDCS (Rosati et
al. 1999). 
{\bf Right:} Apparent angular diameter of a rich cluster (core radius 250
kpc) and a group of galaxies (core radius 100 kpc) as function of redshift,
compared to the PSF of the two ROSAT instruments at the same off--axis
angle (5 armin).}  

\label{fig:numclus}
\end{figure}


\begin{thebibliography}

\bibitem{} Borgani, S., et al. 1999, ApJ, 517, 40.
\bibitem{} Campana, S., et al. 1999, ApJ,524, 423.
\bibitem{} De Grandi, S., et al. 1999, ApJ,513, L17.
\bibitem{} Henry, P. et al. 1992, ApJ, 386, 408.
\bibitem{} Gioia, I.M., et al. 1990, ApJS, 72, 567.
\bibitem{} Lazzati, D., et al. 1999, ApJ, 524, 414.
\bibitem{} Nichol, R.C., et al. 1999, ApJ, 521, L21
\bibitem{} Panzera, M.R., et al. 2001, in preparation.
\bibitem{} Rosati, P. 1995, PhD thesis (University of Rome).
\bibitem{} Rosati, P., et al. 1998, ApJ, 492, L21.
\bibitem{} Rosati, P., et al. 2000, in ``Large-scale structure in the X--ray
Universe'', in press (astro-ph/0001119).
\bibitem{} Vikhlinin, A., et al. 1998a, ApJ, 498, L21.
\bibitem{} Vikhlinin, A., et al. 1998b, ApJ, 502, 558.

\end{thebibliography}
\end{document}